# Search for Gamma Ray Bursts with the ARGO-YBJ Detector in Shower Mode

B. Bartoli[1,2], P. Bernardini[3,4], X. J. Bi[5], Z. Cao[5], S. Catalanotti[1,2], S. Z. Chen[5,*], T. L. Chen[6], S. W. Cui[7], B. Z. Dai[8], A. D'Amone[3,4], Danzenguluobu[6], I. De Mitri[3,4], B. D'Ettorre Piazzoli[1,9], T. Di Girolamo[1,2], G. Di Sciascio[9], C. F. Feng[10], Zhaoyang Feng[5], Zhenyong Feng[11], W. Gao[5], Q. B. Gou[5], Y. Q. Guo[5], H. H. He[5], Haibing Hu[6], Hongbo Hu[5], M. Iacovacci[1,2], R. Iuppa[12,13], H. Y. Jia[11], Labaciren[6], H. J. Li[6], C. Liu[5], J. Liu[8], M. Y. Liu[6], H. Lu[5], L. L. Ma[5], X. H. Ma[5], G. Mancarella[3,4], S. M. Mari[14,15], G. Marsella[3,4], S. Mastroianni[2], P. Montini[16], C. C. Ning[6], L. Perrone[3,4], P. Pistilli[14,15], P. Salvini[17], R. Santonico[9,18], P. R. Shen[5], X. D. Sheng[5], F. Shi[5], A. Surdo[4], Y. H. Tan[5], P. Vallania[19,20], S. Vernetto[19,20], C. Vigorito[20,21], H. Wang[5], C. Y. Wu[5], H. R. Wu[5], L. Xue[10], Q. Y. Yang[8], X. C. Yang[8], Z. G. Yao[5], A. F. Yuan[6], M. Zha[5], H. M. Zhang[5], L. Zhang[5], X. Y. Zhang[10], Y. Zhang[5], J. Zhao[5], Zhaxiciren[6], Zhaxisangzhu[6], X. X. Zhou[11,*], F. R. Zhu[11], and Q. Q. Zhu[5] (The ARGO-YBJ Collaboration)

[1] Dipartimento di Fisica dell'Università di Napoli "Federico II", Complesso Universitario di Monte Sant'Angelo, via Cinthia, I-80126 Napoli, Italy

[2] Istituto Nazionale di Fisica Nucleare, Sezione di Napoli, Complesso Universitario di Monte Sant'Angelo, via Cinthia, I-80126 Napoli, Italy

[3] Dipartimento Matematica e Fisica "Ennio De Giorgi", Università del Salento, via per Arnesano, I-73100 Lecce, Italy

[4] Istituto Nazionale di Fisica Nucleare, Sezione di Lecce, via per Arnesano, I-73100 Lecce, Italy

[5] Key Laboratory of Particle Astrophysics, Institute of High Energy Physics, Chinese Academy of Sciences, P.O. Box 918, 100049 Beijing, China

[6] Tibet University, 850000 Lhasa, Xizang, China

[7] Hebei Normal University, 050024 Shijiazhuang Hebei, China

[8] Yunnan University, 2 North Cuihu Road, 650091 Kunming, Yunnan, China

[9] Istituto Nazionale di Fisica Nucleare, Sezione di Roma Tor Vergata, via della Ricerca Scientifica 1, I-00133 Roma, Italy

[10] Shandong University, 250100 Jinan, Shandong, China

[11] Southwest Jiaotong University, 610031 Chengdu, Sichuan, China

[12] Dipartimento di Fisica dell'Università di Trento, via Sommarive 14, 38123 Povo, Italy.

[13] Trento Institute for Fundamental Physics and Applications, via Sommarive 14, 38123 Povo, Italy

[14] Dipartimento di Fisica dell'Università "Roma Tre", via della Vasca Navale 84, I-00146 Roma, Italy

[15] Istituto Nazionale di Fisica Nucleare, Sezione di Roma Tre, via della Vasca Navale 84, I-00146 Roma, Italy

[16] Dipartimento di Fisica dell'Università di Roma "La Sapienza" and INFN – Sezione di Roma, piazzale Aldo Moro 2, 00185 Roma, Italy

[17] Istituto Nazionale di Fisica Nucleare, Sezione di Pavia, via Bassi 6, I-27100 Pavia, Italy

[18] Dipartimento di Fisica dell'Universit''a di Roma "Tor Vergata," via della Ricerca Scientifica 1, I-00133 Roma, Italy

[19] Osservatorio Astrofisico di Torino dell'Istituto Nazionale di Astrofisica, via P. Giuria 1, I-10125 Torino, Italy

[20] Istituto Nazionale di Fisica Nucleare, Sezione di Torino, via P. Giuria 1, I-10125 Torino, Italy

[21] Dipartimento di Fisica dell'Università di Torino, via P. Giuria 1, I-10125 Torino, Italy

**Abstract**: The ARGO-YBJ detector, located at the Yangbajing Cosmic Ray Laboratory (4300 m a. s. l., Tibet, China), was a "full coverage" air shower array dedicated to gamma ray astronomy and cosmic ray studies. The wide field of view (~ 2 sr) and high duty cycle (> 86%), made ARGO-YBJ suitable to search for short and unexpected gamma ray emissions like gamma ray bursts (GRBs). Between 2007 November 6 and 2013 February 7, 156 satellite-triggered GRBs (24 of them with known redshift) occurred within the ARGO-YBJ field of view (zenith angle $\theta \leqslant 45^{\circ}$). A search for possible emission associated to these GRBs has been made in the two energy ranges 10-100 GeV and 10-1000 GeV. No significant excess has been found in time coincidence with the satellite detections nor in a time window of one hour after the bursts. Taking into account the EBL absorption, upper limits to the energy fluence at 99% of confidence level have been evaluated, with values ranging from ~ $10^{-5}$ erg cm$^{-2}$ to ~$10^{-1}$ erg cm$^{-2}$.

* Corresponding author: zhouxx@swjtu.edu.cn (X. X. Zhou); chensz@ihep.ac.cn (S. Z. Chen)

## 1. Introduction

Gamma ray bursts (GRBs) are the brightest explosion of gamma rays observed so far, occurring at unpredictable times and random directions in the sky. After almost 50 years since their discovery (Klebesadel et al. 1973), GRBs are still one of the most enigmatic astrophysical phenomena. Up to now over 6000 GRBs with gamma rays of energies from keV to MeV have been observed by dedicated satellites as CGRO-BATSE, Beppo SAX, HETE-2, Swift and Fermi-GBM. The results of BATSE (on board the CGRO) showed that the arrival directions of GRBs are highly isotropic, favoring a cosmological origin (Meegan et al. 1992). Thanks to the launch of Beppo SAX, that provided a precise localization of several GRBs, the first redshifts could be measured, proving that GRBs sources are cosmological objects (Metzger et al. 1997). Now more than 470 redshifts have been measured, ranging from $z$ = 0.0085 for GRB980425[a] to the surprising value of $z$ = 8.1 for GRB090423 (Salvaterra et al. 2009). These data confirm that most GRBs indeed originate at cosmological distances and hence are likely the most energetic phenomena ever occurred since the Big Bang.

The time duration of GRBs usually ranges from a few seconds to tens of seconds, but occasionally can be as long as a few tens of minutes or as short as a few milliseconds. According to the $T_{90}$ parameter, defined as the time interval in which 90% (from 5% to 95%) of the GRB photons is released, GRBs are usually classified into long ($T_{90} > 2$ s) and short ($T_{90} \leq 2$ s) bursts (Kouveliotou et al. 1993). Short GRBs are mostly associated to mergers of compact objects while long GRBs are related to collapsars (Rosswog et al. 2003; Bloom et al. 2006; Campana et al. 2006; Larsson et al. 2015; Zhang et al. 2016).

The GRBs photons detected by satellite instruments are mostly in the keV-MeV energy range. Thanks to EGRET (on board the CGRO) and LAT (on board Fermi), both designed to detect photons in the MeV-GeV energy range, 74 GRBs (up to the time of writing, 2017 March 20) have been observed to contain photons of energy above 1 GeV, and for 20 of them photons of energy above 10 GeV have been detected (Hurley et al. 1994; Fermi-LAT GRBs Web site[b]). In particular Fermi-LAT announced the detection of a 95 GeV photon in coincidence with GRB130427A (Ackermann et al. 2014), the highest energy observed so far from a burst. From these observations we know that at least a fraction of GRBs has a high energy tail, reaching 1-10 GeV or even more. At the same time, some theoretical models also predict the emission of high energy photons from GRBs (Ma et al. 2003; Beloborodov et al. 2014; Hascoet et al. 2015).

Though much progress has been achieved from satellite-based experiments and theoretical efforts, lots of basic questions still remain unresolved (Ackermann et al. 2014). Which are the energy source and the acceleration mechanism of GRBs? Is there any GRB originating in our Galaxy? In order to understand the whole process picture, it is important to improve the sample of GRBs with high energy emission and to measure the multi-wavelength energy spectrum. Furthermore, since the flux of high energy photons should be strongly attenuated by the interaction with the extragalactic background light (EBL), the detection of photons with energy above 10 GeV from high redshift sources could be of extreme importance in constraining the EBL theoretical models.

Due to the limited size of the space detectors and the rapid fall of GRB energy spectra, satellite-based experiments hardly cover the energy region $E > 10\text{-}100$ GeV. Ground-based experiments, including imaging atmospheric Cherenkov telescopes (IACTs) and extensive air shower (EAS) arrays, can easily reach much larger effective areas, and can be used for the detection of the GRBs high energy component. Searches for GeV-TeV emission from GRBs have been done by many ground-based experiments, such as MAGIC (Albert et al. 2006; Aleksić et al. 2014), HESS (Abramowski et al. 2014), VERITAS (Acciari et al. 2011), Milagrito (Atkins et al. 2000 & 2005; Abdo et al. 2007), GRAND (Poirier et al. 2003), HEGRA (Padilla et al. 1998), HAWC (Abeysekara et al. 2015), EAS-TOP (Aglietta et al. 1996), INCA (Castellina et al. 2001), Tibet ASγ (Amenomori et al. 1996; Ding et al. 1997; Zhou et al. 2009) and IceCube (Aartsen et al. 2016). However, no significant events have been observed so far, although some positive indications have been reported. The science prospects for GRBs with CTA have been provided by Inoue et al. (2013). The EAS array LHAASO is currently under construction in Sichuan, China, at 4410 m a. s. l. and will begin to take data with a partial array in 2018. The prospects for GRB detection with LHAASO has been discussed by Chen et al. (2015).

Besides a wide field of view and a high duty cycle (fundamental detector properties for the observation of unpredictable and short duration events as GRBs), ARGO-YBJ has two additional key features: the high altitude and the full coverage, making possible the detection of very small showers generated by gamma rays of energy well below 1 TeV. This capability is essential for the GRBs detection, since photons of higher energy originated at cosmological distances are mostly absorbed in the extragalactic space due to pair production with the UV, optical and infrared photons of EBL.

The GRB search by ARGO-YBJ can be done with two different techniques: scaler mode (Vernetto 2000) and shower mode. In scaler mode the data are the single particle counting rate of the detector. A burst candidate would appear as an excess of counts in coincidence with a satellite GRB with no information about the GRB direction. In shower mode, the events are reconstructed and their arrival direction and energy are measured with an offline process, and a burst candidate would appear as a cluster of showers with arrival direction consistent with that of the satellite GRB. The selection of showers according to their direction significantly decreases the background with respect to the scaler mode. On the other hand, the requirement of a minimum number of particles to trigger the detector increases the primary energy threshold.

The ARGO-YBJ results of a search in scaler mode in coincidence with satellite GRBs have been reported in Aielli et al. (2009a) and in Bartoli et al. (2014a) in the energy range 1-100 GeV. In shower mode, the study of the sensitivity to detect GRBs is given by Zhou et al. (2007, 2016), while the results of a first search for GRBs using the data recorded before 2009 January are reported in Aielli et al. (2009b). The results in shower mode concerning the whole sample of data recorded during the ARGO-YBJ lifetime, from 2007 November 6 to 2013 February 7, will be presented and discussed in this paper.

## 2. The ARGO-YBJ experiment

The ARGO-YBJ experiment, located at the Yangbajing Cosmic Ray Laboratory in Tibet, China, at an altitude of 4300 m a. s. l., was mainly devoted to gamma ray astronomy (Bartoli et al. 2013, 2014b, 2015a) and cosmic ray physics (Bartoli et al. 2015b, 2015c). The detector is composed of a single layer of resistive plate chambers (RPCs), operated in streamer mode, with a modular configuration. The basic module is a cluster ($5.7 \times 7.6$ m$^2$), composed of 12 RPCs ($1.23 \times 2.85$ m$^2$ each). Each RPC is read out by 10 pads ($55.6 \times 61.8$ cm$^2$ each), representing the space-time pixels of the detector. The clusters are disposed in a central full-coverage carpet (130 clusters on an area $74 \times 78$ m$^2$ with an active area of ~93%). In order to improve the performance of the experiment in determining the shower core position, the central carpet is surrounded by 23 additional clusters ("guard ring"). The total area of the array is $110 \times 100$ m$^2$. More details of the detector can be found in Aielli et al. (2006). The installation of the whole detector was completed in the spring of 2007, and the data taking started in 2007 November with a trigger rate of ~3.5 kHz. In 2013 February the detector was definitively switched off.

The ARGO-YBJ detector was connected to two independent data acquisition systems, corresponding to the shower and scaler operation modes. In scaler mode, the total counts of each cluster are recorder every 0.5 s. In shower mode, the detector is triggered when at least 20 pads in the central carpet are fired within a time of 420 ns. The information on the arrival time and location of each hit are recorded to reconstruct the shower front shape and the arrival direction. These data are used for gamma astronomy studies and to the search of GRBs in shower mode.

## 3. Data selection and analysis

From 2007 November 6 to 2013 February 7, 188 GRBs occurred in the ARGO-YBJ field of view (zenith angle $\theta \leqslant 45^\circ$). 99 of them were detected by Fermi (selected from the Fermi GBM Burst Catalog Web site[c]) and 89 by Swift and other detectors (selected from Swift and GCN Web sites[d]). The present analysis is carried out on 156 of them, since the remaining 32 occurred when the detector was not operating, or the localization by Fermi was too poor (i.e. error boxes larger than $10^\circ$).

As GRB photons of energy above ~1 TeV (or even below, for very large distances) are likely to be absorbed by the EBL, we assume a GRB power law spectrum with a sharp cutoff at two different maximum energies: 100 GeV and 1 TeV. Since showers generated by photons of energy $E < 10$ GeV cannot trigger the ARGO-YBJ detector, we limit our search to the two energy ranges 10-100 GeV and 10-1000 GeV. The selection of events with these primary energies is done by selecting the showers according to the number of fired pads ($N_{pad}$). To evaluate the $N_{pad}$ intervals optimized for the two energy ranges considered, we use Monte Carlo simulations: the CORSIKA7.3700 code to describe the development of extensive air showers in the atmosphere, and a code based on GEANT4 (Agostinelli et al. 2003) to simulate the detector response. The corresponding intervals are $N_{pad} = 20$-$60$ for $E = 10$-$100$ GeV and $N_{pad} = 20$-$500$ for $E = 10$-$1000$ GeV (see Table 1).

The effective area $A_{eff}$ of ARGO-YBJ in the two $N_{pad}$ intervals has been calculated for gamma ray energies from 10 to 1000 GeV using the expression:

$$A_{eff}(E, \theta) = \frac{n_s}{N} \cdot A_s \cdot \cos \theta \,, \qquad (1)$$

where $n_s$ is the number of successfully reconstructed shower events in a given $N_{pad}$ range, $N$ is the total number of events generated by CORSIKA, and $A_s$ is the sampling area ($200 \times 200$ m$^2$). Fig. 1 and 2 show the effective area as a function of the primary energy in the two $N_{pad}$ ranges, for different zenith angles. The effective area for vertical showers with $N_{pad} = 20$-$60$ ranges from ~0.01 m$^2$ at 10 GeV to ~50 m$^2$ at 100 GeV, while the effective area reaches ~10000 m$^2$ at 1 TeV using the events with $N_{pad} =$ 20-500. The effective areas significantly decrease at large zenith angles. At 100 GeV, the effective area for $\theta = 30°$ (40°) is about a factor 6 (50) smaller than for vertical photons.

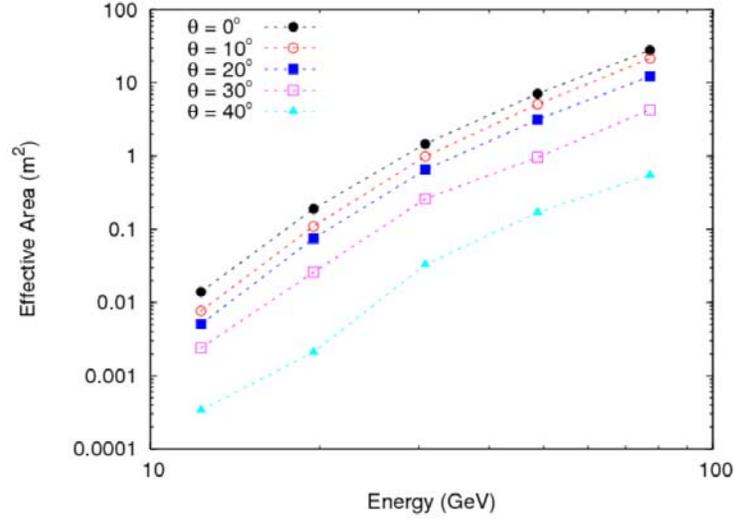

**Figure 1.** Effective area of ARGO-YBJ for events with $N_{pad} = 20$-60 and different zenith angles, as a function of the primary energy.

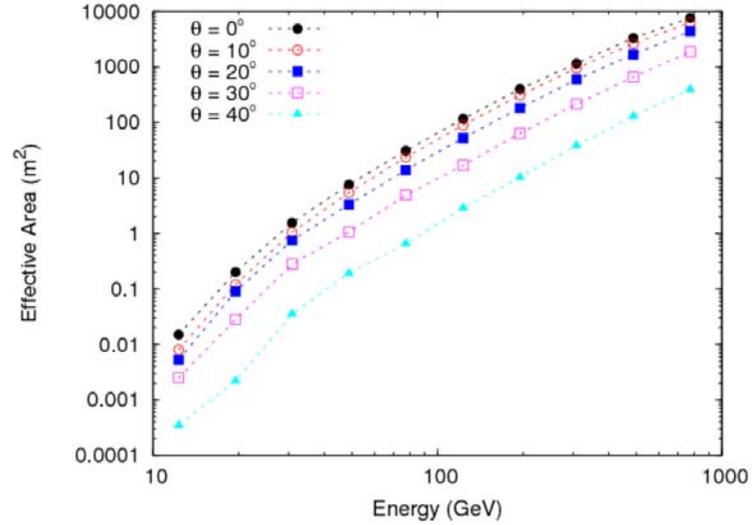

**Figure 2.** Effective area of ARGO-YBJ for events with $N_{pad} = 20$-500 and different zenith angles, as a function of the primary energy.

In the ARGO-YBJ data, a GRB would appear as a cluster of showers with arrival directions concentrated in a small sky window, whose size is related to the angular resolution of the detector. In our analysis, the radius of the search window (defined as the "on-source window") is chosen to maximize the signal to background ratio. According to simulations, the best opening angle radius $\Psi_{70}$ is 3.8° and 2.6° for the two $N_{pad}$ ranges adopted in the analysis, respectively, and contains 71.5% of the signal events.

**Table 1**

$N_{pad}$ ranges and corresponding angular window radii for gamma rays in two different energy ranges

| Energy (GeV) | $N_{pad}$ | $\Psi_{70}$ (°) |
|:---:|:---:|:---:|
| 10-100 | 20-60 | 3.8 |
| 10-1000 | 20-500 | 2.6 |

Due to the different angular resolution to determine the arrival direction of GRBs by different satellite experiments, the position of our on-source window is defined by two approaches. For the 78 GRBs detected by Swift and other satellites excluding Fermi (we call them for simplicity "Swift GRBs" since the large majority of them has been detected by Swift), the position is determined with a precision much better than the angular resolution of ARGO-YBJ. In this case the on-source window is centered at the position reported by the satellite. For 77 of 78 GRBs detected by Fermi-GBM (we call them "Fermi GRBs"), the uncertainty in the position is often greater than 1° (Connaughton et al. 2015). In this case our on-source window is shifted inside the Fermi error box, by steps of $\Psi_{70}/2$ in right ascension and declination, to cover the whole region (defined by including statistical and systematic errors). GRB090902B, even if detected by Fermi, has been analyzed as a "Swift GRB" because of the precise localization determined by Swift.

From the observations of satellite instruments (Hurley et al. 1994; Abdo et al. 2009) and some theoretical models (Dermer & Chiang 2000; Pe'er & Waxman 2004), we know that the acceleration mechanism for high-energy gamma rays could be different from that at low energies, and consequently the burst start time and duration could be different from what recorded in the keV-MeV energy range. The Fermi-LAT results and recent models all show that the high energy emissions from GRBs are in the prompt phase or at delayed times (Hascoet et al. 2015). To take into account these possibilities, our search for GRB counterparts is performed in one hour after the GRB satellite trigger time. To investigate possible different durations of the high energy emission, we use the $T_{90}$ given by satellite measurements, and also the time windows $\Delta t$ = 0.5, 1, 3, 6, 12, 24, 48 and 96 s, that are shifted by steps of 0.25, 0.5, 1, 2, 3, 6, 12 and 24 s, respectively, inside the interval of one hour after the GRB start time.

For each trial, defined by time duration, start time and on-source window position, we compare the number of detected events $N_{on}$ with the number of the expected background events due to cosmic rays $<N_b>$, and we calculate the chance probability $P_b$ of having a number of events equal or larger than $N_{on}$, and the corresponding statistical significance $S$ in standard deviations (s.d.).

The number of the expected background events $<N_b>$ is estimated using the "equi-zenith-angle" method (Zhou 2003), i.e. using the events with the same zenith angle but with different azimuth, detected during two hours around the GRB time. The choice of this time interval comes from the observation that the cosmic ray detection rate (that can fluctuate up to a few percent on time scales of several hours due to atmospheric pressure and temperature variations) in two hours can be considered stable enough for our purposes. Note that due to the atmospheric absorption the background rate is strongly dependent on the zenith angle. As an example, Fig. 3 shows the average background event rate $<N_b>$ as a function of the zenith angle, measured during two hours around the time of GRB110705364, for the two angular windows of radius 3.8° and 2.6°.

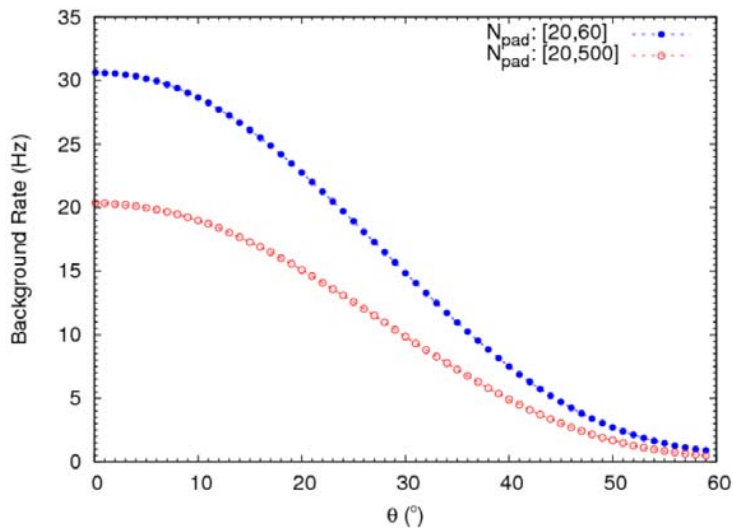

**Figure 3**. Background event rate as a function of the zenith angle, inside the angular windows used for the analysis in two $N_{pad}$ ranges. The data have been recorded during two hours around the GRB110705364 trigger time.

## 4. Results

### 4.1. Analysis of Swift GRBs

The distribution of the chance probability $P_b$ for all the events in coincidence with the 79 Swift GRBs is shown in Fig. 4, while the corresponding significances $S$ are given in Fig. 5. These figures show the results of the analysis in the two $N_{pad}$ ranges previously defined, merged together. The significance distribution is consistent with a Gauss function (solid black line in the figure), showing that the cluster multiplicities are consistent with background fluctuations. It is however interesting to report the details of the most significant clusters.

For the $N_{pad}$ range 20-60, the cluster with the lowest probability is related to GRB080207, whose position $(\alpha_o, \delta_o)$ determined by Swift is $(13^h50^m03^s.12, 07°31'01'')$. It is delayed by 2502 s, and occurs at a zenith angle of 23.5°. This cluster contains 746 showers in 24 s, while the expected number of background events is 617.96. The corresponding poissonian chance probability is $2.98 \times 10^{-7}$, with a significance of 5.0 s.d.. Taking into account for the number of trials $(2.05 \times 10^{6})$, the chance probability is $6.11 \times 10^{-1}$ (-0.28 s.d.).

For the $N_{\text{pad}}$ range 20-500, the lowest probability cluster refers to GRB081025, whose position $(\alpha_{\text{o}}, \delta_{\text{o}})$ determined by Swift is $(16^{\text{h}}21^{\text{m}}11^{\text{s}}.52, 60^{\text{o}}27'58'')$. It is delayed by 3186 s, and occurs at a zenith angle of $32.8^{\text{o}}$. This cluster contains 28 air showers in 1 s, while the expected number of background events is 8.88. The chance probability is $1.53 \times 10^{-7}$, corresponding to 5.12 s.d.. After taking into account for the number of trials, the chance probability is $3.13 \times 10^{-1}$ (0.49 s.d.).

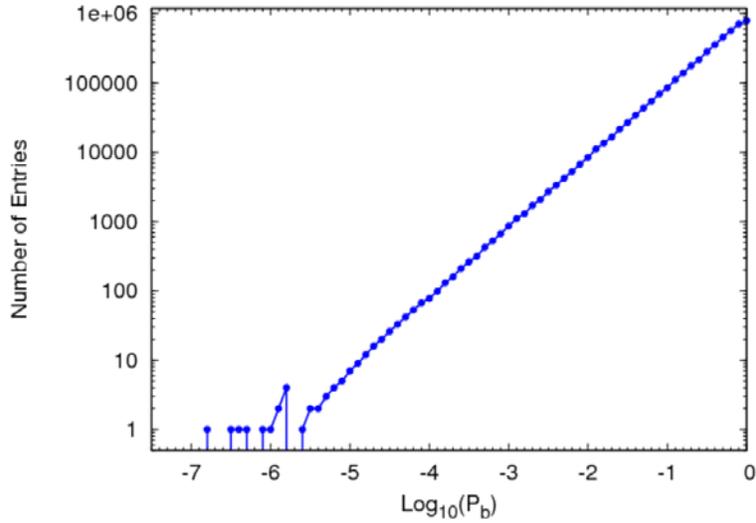

**Figure 4.** Probability distribution of the $4.1 \times 10^{6}$ clusters in coincidence with 79 Swift GRBs.

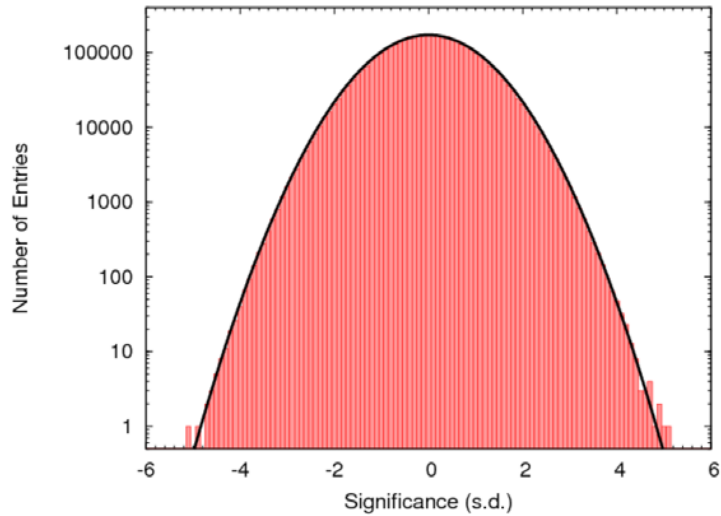

**Figure 5.** Significance distribution of the $4.1 \times 10^{6}$ clusters in coincidence with 79 Swift GRBs. The solid black line is the normal Gaussian function.

## 4.2. Analysis of Fermi GRBs

The distribution of the chance probability $P_{\text{b}}$ for all the clusters in coincidence with 77 Fermi GRBs is shown in Fig. 6, while the corresponding significances $S$ are given in Fig. 7. Also in this case the distributions show no statistically significant excess for this sample.

For the $N_{pad}$ range 20-60, the lowest probability cluster is related to GRB081122520, whose position ($\alpha_o$, $\delta_o$) determined by Fermi is ($22^h36^m24^s$, $40^o00'00''$) with an angular indetermination of 3.8 degrees. The cluster is centered at ($\alpha$, $\delta$) = ($22^h51^m35^s.98$, $40^o00'00''$), at a zenith angle of $10.0^o$. It is delayed by 1176 s and contains 855 air showers in 24 s, while the expected number of background events is 693.4. The chance probability is $1.57 \times 10^{-9}$, corresponding to a significance of 5.92 s.d.. Taking into account for the number of trials ($9.01 \times 10^7$), the chance probability becomes $1.41 \times 10^{-1}$ (1.07 s.d.).

For the $N_{pad}$ range 20-500, the lowest probability cluster is connected to GRB100210101, whose position ($\alpha_o$, $\delta_o$) determined by Fermi is ($16^h17^m31^s.2$, $16^o04'48''$) with an uncertainty of 7.1 degrees. The cluster is centered at ($\alpha$, $\delta$) = ($16^h07^m07^s.18$, $12^o10'48''$), at a zenith angle of $32.1^o$. It is delayed 881 s and contains 64 air showers in 3 s, while the expected number of background events is 27.5. The chance probability is $1.38 \times 10^{-9}$, corresponding to a significance of 5.95 s.d.. After taking into account for the number of trials ($1.79 \times 10^8$), the chance probability is $2.47 \times 10^{-1}$ (0.68 s.d.).

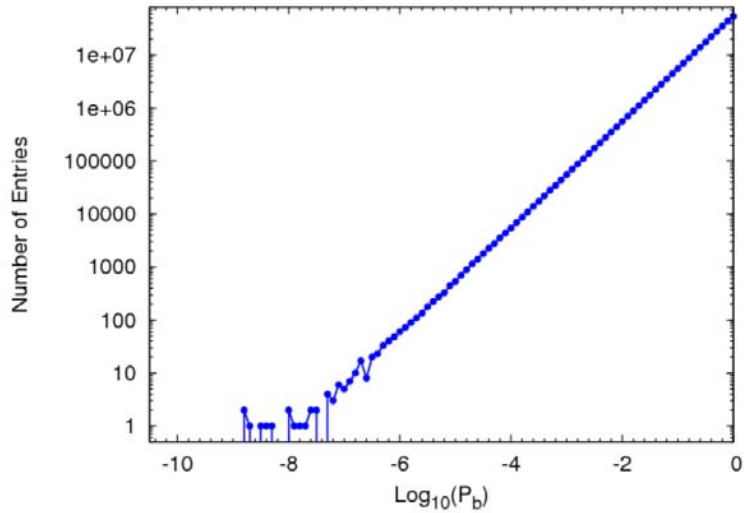

**Figure 6**. Probability distribution of the $2.69 \times 10^8$ clusters in coincidence with 77 Fermi GRBs.

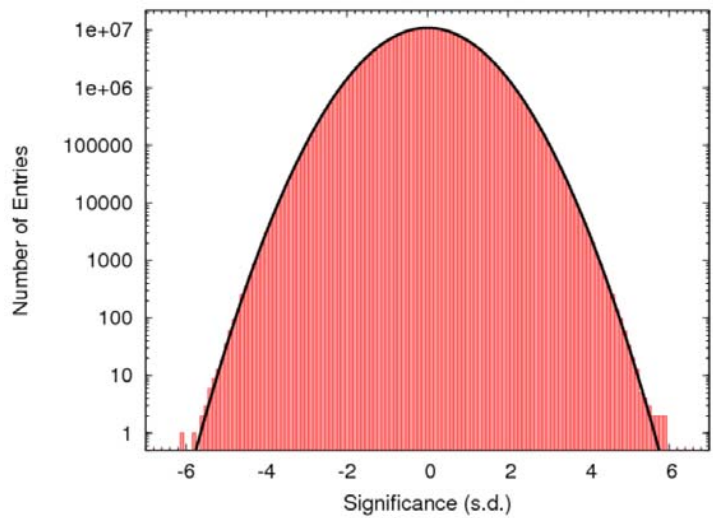

**Figure 7**. Significance distribution of $2.69 \times 10^8$ clusters in coincidence with 77 Fermi GRBs. The solid black line is the normal Gaussian function.

# 5. Upper limits to the energy fluence in $T_{90}$

Since no significant excess has been found in coincidence with 156 satellite-triggered GRBs, we calculate the upper limits to the energy fluence for the two energy ranges considered in the analysis. The fluence upper limits are evaluated during the prompt phase of the emission, i.e. in the time interval $T_{90}$. Fig. 8 and 9 show the significance distributions of the clusters observed in coincidence with 156 GRBs during $T_{90}$ for two $N_{pad}$ ranges, respectively. For Fermi triggered GRBs, the position of the on-source window is set at the center of the error box. No significance larger than 3 standard deviations is observed.

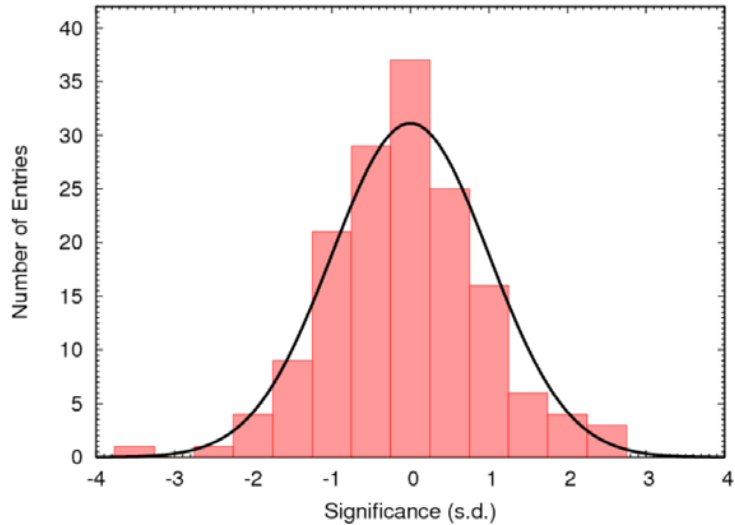

**Figure 8**. Significance distribution of the clusters detected in coincidence with the prompt phase of 156 GRBs, for the $N_{pad}$ range 20-60. The solid line represents the normal Gauss function.

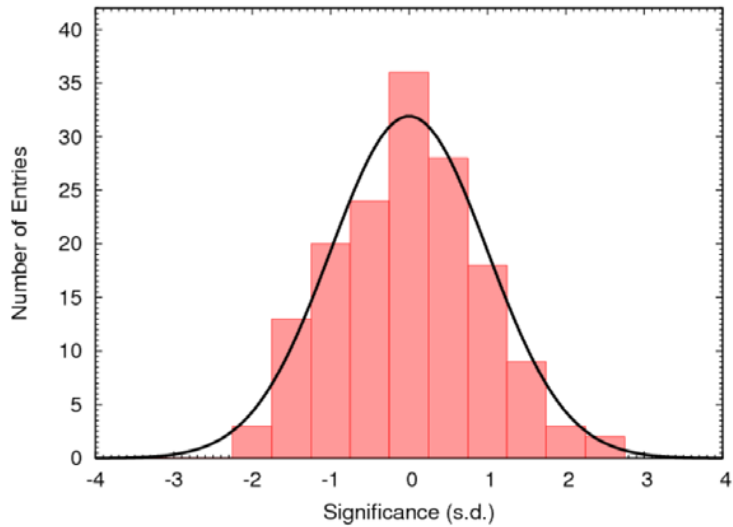

**Figure 9**. Significance distribution of the clusters detected in coincidence with the prompt phase of 156 GRBs, for the $N_{pad}$ range 20-500. The solid line represents a normal Gauss function.

Given a cluster with a number of detected events $N_{on}$ during the interval $T_{90}$ and an expected number of background events $<N_b>$, we calculated the 99% confidence level (C. L.) upper limit on the number of signal events $N_{UL}$, by using the Feldman-Cousins prescription (Feldman & Cousins 1998). To determine the fluence corresponding to the number of events $N_{UL}$, we have to make assumptions on the spectral shape of the intrinsic GRB spectrum at high energy and take into account the absorption due to the EBL. According to the observations in the GeV range by CGRO-EGRET (Dingus 2001) and Fermi-LAT (Ackermann et al. 2013), the spectrum has usually a power law shape and the value of the spectral index at high energy is around 2 ($<\alpha_{LAT}> = 2.05 \pm 0.03$ for 35 GRBs see by Fermi-LAT). We assume an intrinsic power law spectrum $dN/dE = KE^{-\alpha}$ with $\alpha = 2.0$, where the normalization $K$ is obtained from the relation:

$$N_{UL} = K \int_{10\,GeV}^{E_{cut}} A_{eff} \cdot E^{-\alpha} \cdot e^{-\tau_{EBL}} \cdot dE, \qquad (2)$$

Here $\tau_{EBL}$ is the optical depth due to the EBL absorption, and $E_{cut}$ is the maximum energy of the spectrum ($E_{cut} = 100$ GeV or 1 TeV). Concerning the EBL absorption, we use the Gilmore model with a semi-analytical approach (Gilmore et al. 2012), whose parameters at different energies and redshifts can be downloaded from the web site[e]. Finally, the upper limit to the fluence $F_{UL}$ is given by:

$$F_{UL} = K \int_{10\,GeV}^{E_{cut}} E \cdot E^{-\alpha} \cdot dE. \qquad (3)$$

Note that the redshift was measured for only 24 GRBs of our sample. For the bursts with unknown redshift, we calculate the EBL absorption assuming the mean observed redshift $z = 0.6$ for short GRBs ($T_{90} \leq 2$ s) and $z = 2.0$ for long GRBs ($T_{90} > 2$ s) (Jakobsson et al. 2006; Berger et al. 2005; Berger 2014).

Table 2 shows information on the 132 GRBs with unknown redshift, together with the calculated fluence upper limits that are in the range $\sim 10^{-5}$-$10^{-2}$ erg cm$^{-2}$ for $E_{cut} = 100$ GeV and $\sim 10^{-5}$-$10^{-1}$ erg cm$^{-2}$ for $E_{cut} = 1$ TeV.

Fig. 10 and Fig. 11 show the fluence upper limits as a function of the zenith angle for the two energy ranges considered, respectively. The upper limits increase with the zenith angle, due to the showers absorption at larger atmospheric thickness. In both figures it is evident that short GRBs (assumed at $z = 0.6$) have upper limits much smaller than long GRBs (assumed at $z = 2.0$), due to the lower effect of the EBL absorption.

---

[e]  http://physics.ucsc.edu/~joel/EBLdata-Gilmore2012/

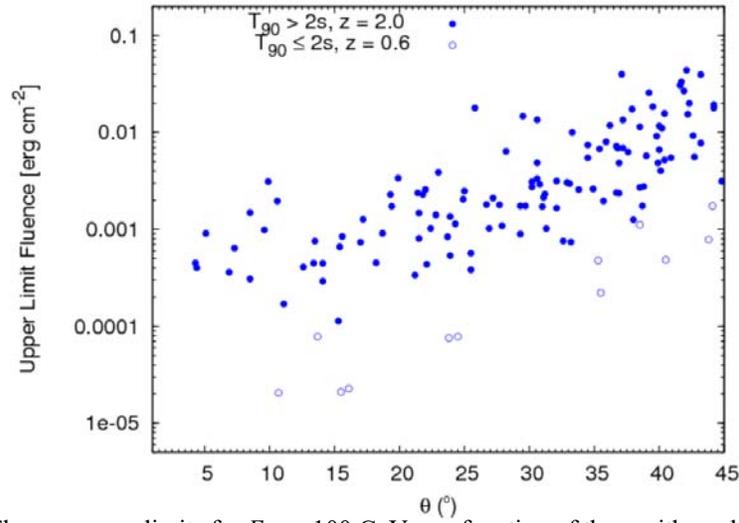

**Figure 10**. Fluence upper limits for $E_{cut}$ = 100 GeV as a function of the zenith angle. Full dots are for long GRBs, circles for short ones.

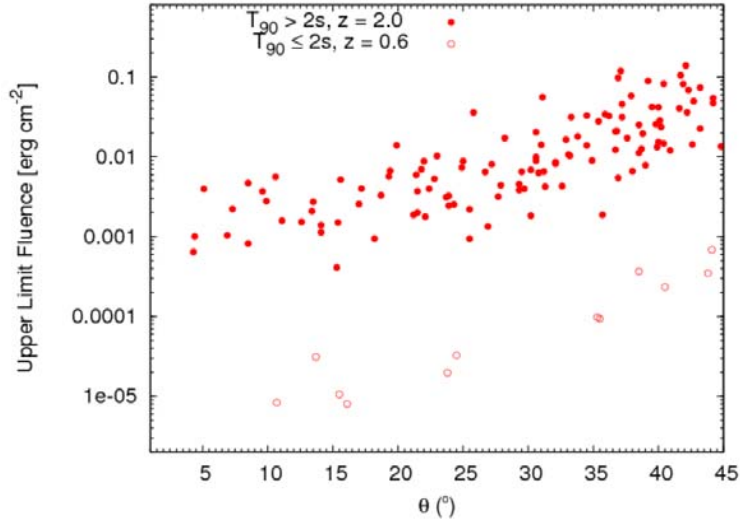

**Figure 11**. Fluence upper limits for $E_{cut}$ = 1 TeV as a function of the zenith angle. Full dots are for long GRBs, circles for short ones.

For the 24 GRBs with known redshift, a more accurate estimation of the absorption can be done, providing more reliable fluence upper limits, given in Table 3 together with the GRBs data. Also in this case the fluence limits are calculated assuming the spectral slope $\alpha$ = 2.0. But for GRB090902B, the spectral slope $\alpha$ =1.94 measured by Fermi-LAT is used.

Fig. 12 shows the $F_{UL}$ values as a function of redshift, for the two energy ranges considered in the analysis. The fluence upper limits increase with redshift, since the flux is more absorbed for far sources. It is interesting to note that, for a given GRB, the fluence for $E_{cut}$ = 100 GeV is higher than that for $E_{cut}$ =1 TeV when $z$ < 1; for $z$ >1 the situation is reversed, due to the large absorption of photons with energy above 100 GeV for $z$ > 1.

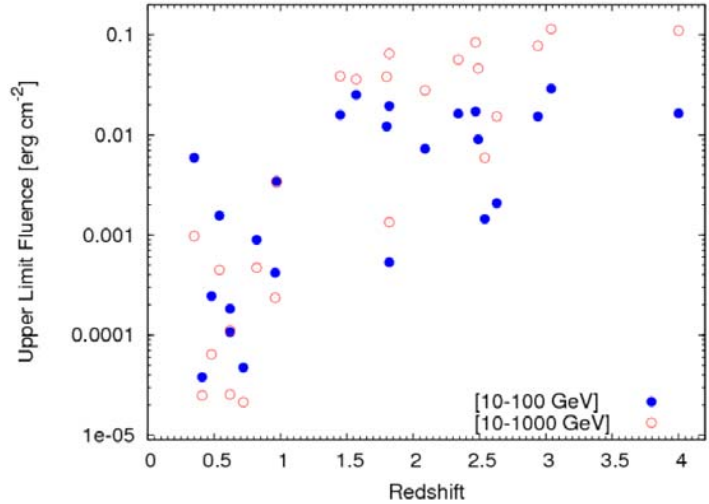

**Figure 12**. Fluence upper limits as a function of redshift.

Among the GRBs occurred in the ARGO-YBJ field of view, GRB090902B is the Fermi-GBM burst with the highest energy photon (33.4 GeV) detected by Fermi-LAT (Abdo et al. 2009), and for this reason it deserves a special attention. The location was precisely determined by Swift, corresponding to a zenith angle of $23.3^{\circ}$ in the ARGO-YBJ field of view. The value of $T_{90}$ is 19.3 s, the redshift 1.82 and the spectral index 1.94. According to our analysis, the clusters observed by ARGO-YBJ at the burst trigger time inside the interval $T_{90}$ for the two $N_{pad}$ ranges (20-60) and (20-500) have a significance of -0.89 and -1.02 s.d., respectively. Since the 33.4 GeV photon was detected 82 seconds after the GBM trigger, i.e. about a minute after the end of the prompt emission, we did a search in coincidence with this delayed GeV emission. We considered 3 time windows: *a)* the interval (0-90 s), including the most energetic photon, *b)* the interval (6-26 s), when the maximum density of GeV photons has been observed, *c)* the interval (82-83 s), i.e. the time around the 33.4 GeV photon. The significance of the clusters detected in the 3 time windows are -1.53, -1.05 and -0.07 s.d., respectively, for $N_{pad} = 20$-60, and -0.45, -0.41 and 0.74 s.d., respectively, for $N_{pad} = 20$-500. Using the set of time windows for the standard analysis, the maximum significance is 4.33 s.d. for a time interval of 3 s delayed 2047 s since the burst trigger time. After taking into account for the number of trials, the significance lowers to 0.22 s.d. leading to the conclusion that there are no significant clusters associated to this GRB. Following the method described at the beginning of this section, we evaluated the fluence upper limit in $T_{90}$ at the prompt phase. We found $F_{UL} = 5.32 \times 10^{-4}$ erg cm$^{-2}$ ($1.74 \times 10^{-4}$ erg cm$^{-2}$ without correction for the EBL absorption) in the energy range 10-100 GeV to be compared with the extrapolated fluence $F_{EX} = 1.06 \times 10^{-4}$ erg cm$^{-2}$ (obtained by extrapolating up to $E_{cut}$ the flux measured by satellite). In the energy range 10-1000 GeV, we found $F_{UL} = 1.34 \times 10^{-3}$ erg $cm^{-2}$ ($4.63 \times 10^{-6}$ erg $cm^{-2}$ without correction for EBL absorption) to be compared with $F_{EX} = 2.28 \times 10^{-4}$ erg $cm^{-2}$. Taking into account the EBL absorption, the extrapolated fluences of GRB090902B are lower than the upper limits measured by ARGO-YBJ. According to our calculations, ARGO-YBJ could detect GRB090902B for a redshift lesser than 1.0 and an emission extending up to 1 TeV.

## 6. Detectability of GRB 130427A

GRB 130427A, detected by both Fermi-GBM and Fermi-LAT, is the burst with the highest energy photons ever observed, reaching 95 GeV. ARGO-YBJ was switched off in 2013 February, two months before the burst, but since the zenith angle of the GRB was greater than 90°, the burst would have been undetectable in any case. It is interesting however to estimate the capability of ARGO-YBJ to detect a burst as bright as GRB 130427A, in case of a more favorable zenith angle.

The measured spectral index of GRB 130427A ranges between $\alpha \sim 2.5$ and $\alpha \sim 1.7$, the total fluence in the 10 keV-100 GeV band is $4.9 \times 10^{-3}$ erg $cm^{-2}$ and the duration is $T_{90} = 138.2$ s (Ackermann et al. 2014). Assuming a zenith angle $\theta = 20°$, a spectral index $\alpha = 2.0$, and considering the EBL absorption corresponding to the GRB distance ($z = 0.34$), the estimated minimum fluence detectable by ARGO-YBJ in $T_{90}$ is $1.27 \times 10^{-3}$ erg $cm^{-2}$ in the energy range 10-100 GeV, and $2.38 \times 10^{-4}$ erg $cm^{-2}$ in the energy range 10-1000 GeV. Assuming a spectrum extending up to 1 TeV with the same spectral index, the fluence of GRB 130427A in the energy range 10-1000 GeV would be $4.7 \times 10^{-4}$ erg $cm^{-2}$, twice higher than the ARGO-YBJ sensitivity. Thus, an overhead burst as bright as GRB 130427A and with a photon emission extending up to 1 TeV would have been easily detected by ARGO-YBJ.

## 7. Conclusions

A search for GeV-TeV burst-like events in coincidence with GRBs detected by satellite was done using the ARGO-YBJ data from 2007 November 6 to 2013 February 7. During more than five years, a total of 156 GRBs was analyzed. After considering the number of trials, the distribution of the chance probabilities for the clusters of events occurred in coincidence with the GRBs (i.e. during $T_{90}$) or in a time window of one hour after the satellite trigger time is consistent with background fluctuations. None of the examined GRBs, notably the 'Swift GRBs', contains high energy (> 10 GeV) photons with fluxes comparable to that featuring GRB 130427A and spectrum extending up to 1 TeV. Indeed many conditions, as the efficiency of the production mechanism, the source transparency and the attenuation by the EBL, may affect the GeV-TeV fluence .

Since no significant signal has been detected, the 99% C. L. upper limits to the energy fluence were evaluated. The limits are calculated assuming a power law energy spectrum with spectral index $\alpha$ = 2.0 above 10 GeV, a sharp cutoff at the $E_{cut}$ energy (100 GeV or 1 TeV), and taking into account the EBL absorption (assuming a redshift $z$ = 0.6 for short GRBs and $z$ = 2.0 for long GRBs when the redshift was not available). The obtained fluence upper limits cover a large range of values, depending on the GRB duration, zenith angle and distance, ranging from less than $10^{-5}$ erg cm$^{-2}$ to $10^{-1}$ erg cm$^{-2}$.

In the next future, HAWC, CTA and LHAASO (these two currently under construction), given their significantly higher sensitivity, are expected to provide the first measurement of the GeV-TeV emission from gamma ray bursts by ground-based detectors.

### Acknowledgements

This work is supported in China by the National Natural Science Foundation of China (NSFC) under the Grant Nos. 11475141, 11575203 and 11375209, the Chinese Ministry of Science and Technology, the Chinese Academy of Sciences (CAS), the Key laboratory of Particle Astrophysics, Institute of High Energy Physics (IHEP), and in Italy by the Istituto Nazionale di Fisica Nucleare (INFN).

**Table 2**

List on 132 GRBs with unknown redshift occurred in the field of view of ARGO-YBJ

| GRB (1) | Satellite (2) | $T_{90}$ (s) (3) | $\theta$ (°) (4) | keV fluence ($10^{-7}$erg cm$^{-2}$) (keV range) (5) | $\sigma_1$ (10-100 GeV) (6) | $\sigma_2$ (10-1000 GeV) (7) | $F_{UL1}$ (erg cm$^{-2}$) (10-100 GeV) (8) | $F_{UL2}$ (erg cm$^{-2}$) (10-1000 GeV) (9) |
|---|---|---|---|---|---|---|---|---|
| 080328 | Swift | 90.6 | 37.2 | 94 (15-150) | -0.04 | -1.69 | 1.35E-2 | 4.56E-2 |
| 080515 | Swift | 21 | 43.2 | 20 (15-150) | -0.59 | -0.41 | 7.77E-3 | 2.26E-2 |
| 080613B | Swift | 105 | 39.2 | 58 (15-150) | 0.89 | 1.32 | 2.57E-2 | 8.94E-2 |
| 080714086 | Fermi | 5.4 | 24.3 | 6.8 (10-1000) | 1.82 | 0.71 | 1.14E-3 | 2.53E-3 |
| 080726 | AGILE | 12 | 36.7 | … | 1.50 | 1.03 | 7.20E-3 | 2.09E-2 |
| 080727C | Swift | 79.7 | 34.5 | 52 (15-150) | -0.79 | 0.80 | 5.46E-3 | 3.29E-2 |
| 080730520 | Fermi | 17.4 | 31.2 | 48.7 (10-1000) | 0.03 | -0.20 | 2.31E-3 | 6.54E-3 |
| 080802386 | Fermi | 0.6 | 23.8 | 3.98 (10-1000) | -0.01 | -0.70 | 7.57E-5 | 1.97E-5 |
| 080822B | Swift | 64 | 40.4 | 1.7 (15-150) | -1.11 | -2.00 | 1.57E-2 | 8.21E-2 |
| 080830368 | Fermi | 40.9 | 35.9 | 70 (10-1000) | 0.41 | 1.36 | 7.99E-3 | 3.41E-2 |
| 080903 | Swift | 66 | 21.4 | 14 (15-150) | 0.94 | 0.30 | 2.37E-3 | 5.94E-3 |
| 081025 | Swift | 23 | 30.6 | 19 (15-150) | 0.84 | 0.69 | 3.33E-3 | 9.91E-3 |
| 081102365 | Fermi | 1.7 | 35.3 | 10.9 (10-1000) | -0.21 | -1.01 | 4.77E-4 | 9.76E-5 |
| 081105 | IPN | 10 | 36.7 | ... | -0.90 | -1.06 | 2.39E-3 | 1.22E-2 |
| 081122520 | Fermi | 23.3 | 9.9 | 75.4 (10-1000) | 0.78 | 1.35 | 3.11E-3 | 2.79E-3 |
| 081130629 | Fermi | 45.6 | 37.9 | 32.2 (10-1000) | 1.82 | 1.85 | 1.74E-2 | 5.80E-2 |
| 081215784 | Fermi | 5.6 | 31.3 | 547 (10-1000) | -0.56 | 0.09 | 1.02E-3 | 4.24E-3 |
| 081226156 | Fermi | 65.8 | 42.7 | 39.5 (10-1000) | -1.32 | 0.29 | 5.60E-3 | 4.97E-2 |
| 090107A | Swift | 12.2 | 40.1 | 2.3 (15-150) | -0.99 | -1.17 | 4.01E-3 | 2.84E-2 |
| 090118 | Swift | 16 | 13.4 | 4 (15-150) | -0.30 | 0.56 | 4.49E-4 | 2.09E-3 |
| 090301A | Swift | 41 | 14.1 | 230 (15-150) | -1.25 | -1.50 | 4.46E-4 | 1.13E-3 |
| 090301315 | Fermi | 23.3 | 22.8 | 22.7 (10-1000) | 0.56 | 1.20 | 1.41E-3 | 5.30E-3 |
| 090306B | Swift | 20.4 | 38.5 | 31 (15-150) | 1.24 | 0.11 | 1.14E-2 | 2.51E-2 |

* Corresponding author: zhouxx@swjtu.edu.cn (X. X. Zhou); chensz@ihep.ac.cn (S. Z. Chen)

| | | | | | | | | |
|---|---|---|---|---|---|---|---|---|
| 090320801 | Fermi | 29.2 | 22.4 | 16.7 (10-1000) | -0.33 | 0.20 | 1.02E-3 | 3.97E-3 |
| 090328713 | Fermi | 0.2 | 15.5 | 1.19 (10-1000) | -0.60 | -0.27 | 2.10E-5 | 1.06E-5 |
| 090403314 | Fermi | 14.8 | 29.7 | 10.9 (10-1000) | -0.04 | -0.64 | 1.74E-3 | 3.97E-3 |
| 090425377 | Fermi | 75.4 | 44.2 | 181 (10-1000) | -0.22 | 0.31 | 1.79E-2 | 5.41E-2 |
| 090511684 | Fermi | 7.6 | 39.0 | 24.9 (10-1000) | 0.34 | -1.18 | 5.74E-3 | 7.84E-3 |
| 090520A | Swift | 20 | 42.2 | 3.4 (15-150) | 1.27 | 0.67 | 1.54E-2 | 3.60E-2 |
| 090529564 | Fermi | 9.9 | 22.1 | 86.9 (10-1000) | -0.86 | -0.42 | 4.36E-4 | 1.77E-3 |
| 090617208 | Fermi | 0.2 | 16.1 | 9.43 (10-1000) | -0.51 | -0.77 | 2.27E-5 | 8.00E-6 |
| 090621B | Swift | 0.14 | 40.5 | 0.7 (15-150) | 0.20 | 0.40 | 4.84E-4 | 2.34E-4 |
| 090704783 | Fermi | 19.5 | 4.3 | 15.8 (10-1000) | 0.13 | -1.30 | 4.50E-4 | 6.41E-4 |
| 090712 | Swift | 145 | 10.6 | 40 (15-150) | 0.93 | 0.47 | 1.96E-3 | 5.60E-3 |
| 090730608 | Fermi | 9.1 | 4.4 | 31.8 (10-1000) | 0.82 | -0.03 | 4.01E-4 | 1.01E-3 |
| 090802235 | Fermi | 0.1 | 35.5 | 11.4 (10-1000) | 0.07 | 0.43 | 2.21E-4 | 9.37E-5 |
| 090807A | Swift | 140.8 | 19.9 | 22 (15-150) | 1.21 | 2.42 | 3.37E-3 | 1.39E-2 |
| 090807832 | Fermi | 17.9 | 29.3 | 13.4 (10-1000) | -0.07 | -0.44 | 1.74E-3 | 4.56E-3 |
| 090820027 | Fermi | 12.4 | 17.0 | 1540 (10-1000) | 0.80 | 1.09 | 7.37E-4 | 2.56E-3 |
| 090831317 | Fermi | 39.4 | 35.7 | 94.4 (10-1000) | -1.61 | -2.11 | 1.96E-3 | 1.87E-3 |
| 090904A | Swift | 122 | 22.0 | 30 (15-150) | 0.16 | 0.45 | 2.57E-3 | 8.81E-3 |
| 090904581 | Fermi | 38.4 | 32.9 | 16.4 (10-1000) | -0.64 | 0.52 | 3.04E-3 | 1.64E-2 |
| 091030828 | Fermi | 98.1 | 25.0 | 296 (10-1000) | 0.08 | 0.36 | 2.49E-3 | 8.84E-3 |
| 091106762 | Fermi | 14.6 | 30.2 | 18.3 (10-1000) | 1.53 | 0.44 | 3.10E-3 | 6.86E-3 |
| 091202 | Integral | 45 | 33.2 | … | -1.96 | -0.72 | 7.40E-4 | 1.03E-2 |
| 091215234 | Fermi | 4.4 | 25.5 | 9.87 (10-1000) | -0.89 | 0.29 | 3.83E-4 | 2.20E-3 |
| 091227294 | Fermi | 21.9 | 27.9 | 68.9 (10-1000) | -0.86 | -0.40 | 1.09E-3 | 4.40E-3 |
| 100111A | Swift | 12.9 | 21.5 | 6.7 (15-150) | 0.22 | -0.28 | 8.06E-4 | 2.00E-3 |
| 100115A | Swift | 3 | 32.6 | … | -0.85 | 0.19 | 7.57E-4 | 4.30E-3 |
| 100122616 | Fermi | 22.5 | 33.1 | 120 (10-1000) | -0.25 | 0.01 | 2.96E-3 | 1.06E-2 |
| 100131730 | Fermi | 3.5 | 14.1 | 73.4 (10-1000) | 0.29 | 1.43 | 2.91E-4 | 1.39E-3 |
| 100210101 | Fermi | 29.2 | 24.9 | 21.1 (10-1000) | 0.89 | 1.44 | 2.04E-3 | 7.31E-3 |
| 100225703 | Fermi | 4.5 | 8.5 | 16.1 (10-1000) | 0.76 | 0.11 | 3.07E-4 | 8.19E-4 |

| 100513879 | Fermi | 11.1 | 38.7 | 37.1 (10-1000) | -1.57 | -0.71 | 1.74E-3 | 1.26E-2 |
|-----------|-------|------|------|----------------|-------|-------|---------|---------|
| 100522A | Swift | 35.3 | 27.7 | 21 (15-150) | -0.30 | -1.15 | 1.79E-3 | 3.17E-3 |
| 100525744 | Fermi | 1.5 | 13.7 | 6.44 (10-1000) | 0.74 | 0.68 | 7.83E-5 | 3.10E-5 |
| 100526A | Swift | 102 | 9.6 | 25 (15-150) | -0.33 | -0.07 | 9.87E-4 | 3.70E-3 |
| 100527795 | Fermi | 184.6 | 33.3 | 139 (10-1000) | -1.93 | -1.89 | 1.00E-2 | 3.13E-2 |
| 100614498 | Fermi | 172.3 | 43.2 | 196 (10-1000) | -0.16 | -0.80 | 3.96E-2 | 7.36E-2 |
| 100625891 | Fermi | 29.2 | 15.4 | 14 (10-1000) | -0.43 | -0.88 | 6.60E-4 | 1.50E-3 |
| 100713A | Integral | 20 | 12.6 | 210 (20-200) | -0.56 | -0.30 | 4.09E-4 | 1.53E-3 |
| 100714686 | Fermi | 5.6 | 39.9 | 15.6 (10-1000) | 0.08 | -0.24 | 4.84E-3 | 1.32E-2 |
| 100902A | Swift | 428.8 | 37.1 | 32 (15-150) | 1.08 | 0.75 | 3.99E-2 | 1.19E-1 |
| 101003244 | Fermi | 10 | 30.8 | 22.3 (10-1000) | 1.71 | 0.51 | 2.91E-3 | 6.30E-3 |
| 101101744 | Fermi | 3.3 | 25.5 | 6.5 (10-1000) | 0.07 | -1.02 | 5.67E-4 | 9.41E-4 |
| 101107011 | Fermi | 375.8 | 25.8 | 72.6 (10-1000) | 2.55 | 0.51 | 1.79E-2 | 3.59E-2 |
| 101112984 | Fermi | 82.9 | 39.8 | 85.7 (10-1000) | -0.71 | -0.89 | 9.16E-3 | 2.56E-2 |
| 101123952 | Fermi | 103.9 | 28.2 | 1130 (10-1000) | 1.49 | 0.99 | 6.39E-3 | 1.71E-2 |
| 101202154 | Fermi | 18.4 | 38.0 | 14.1 (10-1000) | -1.85 | -1.58 | 1.26E-3 | 6.61E-3 |
| 101231067 | Fermi | 23.6 | 23.9 | 168 (10-1000) | 0.32 | -0.24 | 1.36E-3 | 3.24E-3 |
| 110106A | Swift | 4.3 | 34.9 | 3 (15-150) | 0.65 | 0.77 | 2.61E-3 | 9.01E-3 |
| 110210A | Swift | 233 | 23.0 | 9.6 (15-150) | 0.19 | -0.16 | 3.87E-3 | 1.02E-2 |
| 110220761 | Fermi | 33 | 31.0 | 21.1 (10-1000) | -0.99 | 1.17 | 1.71E-3 | 1.41E-2 |
| 110226989 | Fermi | 14.1 | 36.9 | 19 (10-1000) | -1.16 | -0.86 | 2.36E-3 | 9.74E-2 |
| 110312A | Swift | 28.7 | 37.2 | 8.2 (15-150) | -0.002 | 0.84 | 6.91E-3 | 3.13E-2 |
| 110315A | Swift | 77 | 19.4 | 41 (15-150) | 0.26 | 0.96 | 1.73E-3 | 6.66E-3 |
| 110328520 | Fermi | 141.3 | 17.2 | 190 (10-1000) | -0.97 | -0.96 | 1.27E-3 | 4.00E-3 |
| 110401920 | Fermi | 2.4 | 15.3 | 15.7 (10-1000) | -1.15 | -1.00 | 1.13E-4 | 4.11E-4 |
| 110406A | Integral | 8 | 31.1 | … | 0.85 | 0.35 | 2.13E-3 | 5.60E-2 |
| 110414A | Swift | 152 | 44.2 | 35 (15-150) | -0.84 | -0.69 | 1.93E-2 | 4.69E-2 |
| 110517573 | Fermi | 23 | 29.5 | 87.4 (10-1000) | -0.65 | 0.06 | 1.47E-2 | 6.50E-3 |
| 110605183 | Fermi | 82.7 | 33.8 | 193 (10-1000) | -1.30 | -0.24 | 2.57E-3 | 1.80E-2 |
| 110625A | Swift | 44.5 | 40.0 | 280 (15-150) | -1.07 | -1.39 | 6.61E-3 | 1.53E-2 |

| 110626448 | Fermi | 6.4 | 40.4 | 11.6 (10-1000) | -0.003 | -0.22 | 5.20E-3 | 1.47E-2 |
|---|---|---|---|---|---|---|---|---|
| 110629174 | Fermi | 61.7 | 5.1 | 24.3 (10-1000) | 0.40 | 1.18 | 9.09E-4 | 3.96E-3 |
| 110705364 | Fermi | 19.2 | 18.7 | 89.4 (10-1000) | 0.56 | 0.99 | 9.10E-4 | 3.30E-3 |
| 110706202 | Fermi | 12 | 26.9 | 32.7 (10-1000) | -0.22 | -1.41 | 1.02E-3 | 1.34E-3 |
| 110706977 | Fermi | 33.2 | 29.3 | 65.5 (10-1000) | -1.41 | -1.11 | 8.93E-4 | 3.82E-3 |
| 110709A | Swift | 44.7 | 13.5 | 100 (15-150) | -0.35 | -0.11 | 7.56E-4 | 2.73E-3 |
| 110709463 | Fermi | 24.1 | 26.7 | 69.1 (10-1000) | 0.34 | 0.79 | 1.80E-3 | 6.47E-3 |
| 110820A | Swift | 256 | 41.7 | 8.2 (15-150) | -0.07 | 0.16 | 3.31E-2 | 1.05E-1 |
| 110915A | Swift | 78.76 | 39.5 | 57 (15-150) | 0.42 | -0.34 | 1.84E-2 | 4.20E-2 |
| 110919634 | Fermi | 35.07 | 42.6 | 268 (10-1000) | -0.63 | -1.36 | 9.27E-3 | 1.43E-2 |
| 110921A | Swift | 48 | 7.3 | 24 (15-150) | -0.30 | -0.25 | 6.40E-4 | 2.21E-3 |
| 110928180 | Fermi | 148.2 | 8.5 | 142 (10-1000) | 0.45 | 0.28 | 1.49E-3 | 4.69E-3 |
| 111017657 | Fermi | 11.1 | 40.0 | 207 (10-1000) | 1.78 | 2.28 | 1.16E-2 | 4.17E-2 |
| 111024A | MAXI | 0.2 | 10.7 | … | -0.20 | -0.30 | 2.06E-5 | 8.33E-6 |
| 111103B | Swift | 167 | 41.6 | 80 (15-150) | 0.23 | -0.22 | 3.06E-2 | 4.04E-2 |
| 111109873 | Fermi | 9.7 | 32.1 | 66.9 (10-1000) | -0.26 | 0.81 | 1.66E-3 | 8.16E-3 |
| 111117A | Swift | 0.47 | 38.5 | 1.4 (15-150) | 2.18 | 1.80 | 1.12E-3 | 3.66E-4 |
| 111127810 | Fermi | 19 | 30.2 | 86.4 (10-1000) | 0.81 | -1.63 | 2.74E-3 | 1.83E-3 |
| 111208A | Swift | 20 | 11.1 | 9.8 (15-150) | -1.54 | -0.12 | 1.70E-4 | 1.59E-3 |
| 111215A | Swift | 796 | 30.6 | 45 (15-150) | -3.51 | -1.21 | 1.35E-2 | 2.04E-2 |
| 111228453 | Fermi | 2.9 | 23.9 | 27.5 (10-1000) | 0.479 | 1.68 | 5.37E-4 | 2.44E-3 |
| 120102A | Swift | 38.7 | 44.8 | 43 (15-150) | -1.80 | -1.36 | 3.14E-3 | 1.34E-2 |
| 120106A | Swift | 61.6 | 35.4 | 9.7 (15-150) | -0.27 | 0.26 | 6.76E-3 | 2.77E-2 |
| 120118898 | Fermi | 17.2 | 18.2 | 16.2 (10-1000) | -0.75 | -1.26 | 4.53E-4 | 9.46E-4 |
| 120129A | IPN | 4 | 38.5 | … | -0.44 | -0.01 | 2.71E-3 | 1.12E-2 |
| 120202A | Integral | 100 | 15.6 | 0.7 (20-200) | -0.98 | 0.38 | 8.43E-4 | 5.17E-3 |
| 120217808 | Fermi | 5.9 | 38.8 | 17.5 (10-1000) | -0.74 | 1.01 | 2.77E-3 | 1.96E-2 |
| 120219A | Swift | 90.5 | 32.1 | 5.4 (15-150) | -1.09 | -1.25 | 3.16E-3 | 8.56E-3 |
| 120222021 | Fermi | 1.1 | 44.1 | 17.3 (10-1000) | 2.39 | 1.72 | 1.74E-3 | 6.81E-4 |
| 120223933 | Fermi | 14.3 | 37.6 | 38.8 (10-1000) | 0.46 | 0.003 | 6.28E-3 | 1.71E-2 |

| 120226447 | Fermi | 14.6 | 36.9 | 58.5 (10-1000) | -0.003 | -1.51 | 4.83E-3 | 5.44E-3 |
|---|---|---|---|---|---|---|---|---|
| 120512A | Integral | 40 | 36.8 | ... | -2.52 | -0.34 | 6.90E-3 | 2.10E-2 |
| 120519721 | IPN | 0.96 | 43.8 | 24.1 (15-150) | 0.06 | -0.05 | 7.83E-4 | 3.48E-4 |
| 120522361 | Fermi | 28.2 | 40.2 | 93.2 (10-1000) | 0.17 | -0.59 | 1.10E-2 | 2.37E-2 |
| 120612680 | Fermi | 63.2 | 21.5 | 20.6 (10-1000) | -0.31 | -0.70 | 1.47E-3 | 3.71E-3 |
| 120625119 | Fermi | 7.4 | 21.2 | 102 (10-1000) | -0.89 | 0.21 | 3.37E-4 | 1.87E-3 |
| 120703498 | Fermi | 77.6 | 21.8 | 26 (10-1000) | 0.49 | 0.47 | 2.29E-3 | 6.97E-3 |
| 120819A | Swift | 71 | 42.3 | 14 (15-150) | 0.29 | 0.91 | 2.00E-2 | 6.87E-2 |
| 120905657 | Fermi | 195.6 | 41.9 | 195.7 (10-1000) | -0.30 | -0.15 | 2.67E-2 | 8.16E-2 |
| 120915474 | Fermi | 5.9 | 40.9 | 3.8 (10-1000) | 0.08 | -0.57 | 5.47E-3 | 1.20E-2 |
| 121011A | Swift | 75.6 | 19.3 | 27 (15-150) | 1.10 | 0.38 | 2.29E-3 | 5.66E-3 |
| 121012724 | Fermi | 0.45 | 24.5 | 12.7 (10-1000) | 0.19 | 0.57 | 7.83E-5 | 3.26E-5 |
| 121025A | MAXI | 20 | 6.9 | ... | -0.42 | -0.72 | 3.61E-4 | 1.04E-3 |
| 121108A | Swift | 89 | 36.2 | 9.6 (15-150) | 0.19 | -0.10 | 1.18E-2 | 3.24E-2 |
| 121113544 | Fermi | 95.5 | 34.5 | 268.5 (10-1000) | -0.32 | -1.07 | 7.43E-3 | 1.39E-2 |
| 121123A | Swift | 317 | 42.1 | 150 (15-150) | 0.25 | 0.60 | 4.38E-2 | 1.38E-1 |
| 121202A | Swift | 20.1 | 27.2 | 12 (15-150) | 0.99 | 1.81 | 2.10E-3 | 8.04E-3 |
| 121211695 | Fermi | 9.0 | 23.7 | 13.4 (10-1000) | 0.31 | 0.83 | 8.39E-4 | 3.11E-3 |
| 130122A | Swift | 64 | 30.6 | 7.4 (15-150) | 0.48 | -0.53 | 4.85E-3 | 8.94E-3 |

**Notes**. Column 1: GRB name (GCN name for Swift GRBs and Trigger ID for Fermi GRBs). Column 2: satellite that detected the GRB. Column 3: burst duration $T_{90}$ as measured by the respective satellite. Column 4: zenith angle at the ARGO-YBJ location. Column 5: fluence in the keV range measured by the satellite that detected the burst. Column 6 and 7: statistical significance of the ARGO-YBJ cluster detected during the prompt phase, for two values of $E_{cut}$ (100 GeV and 1 TeV). Column 8 and 9: 99% C. L. upper limits to the fluence for two values of $E_{cut}$ (100 GeV and 1 TeV).

**Table 3**

List of 24 GRBs with known redshift occurred in the field of view of ARGO-YBJ

| GRB (1) | Satellite (2) | $\alpha_{SAT}$ (3) | $z_{SAT}$ (4) | $T_{90}$ (s) (5) | $\theta$ (°) (6) | keV fluence ($10^{-7}$erg cm$^{-2}$) (keV range) (7) | $\sigma_1$ (10-100 GeV) (8) | $\sigma_2$ (10-1000 GeV) (9) | $F_{UL1}$ (erg cm$^{-2}$) (10-100 GeV) (10) | $F_{UL2}$ (erg cm$^{-2}$) (10-1000 GeV) (11) |
|---|---|---|---|---|---|---|---|---|---|---|
| 071112C | Swift | 1.09 | 0.82 | 15 | 18.42 | 30 (15-150) | 0.25 | 0.12 | 8.90E-4 | 4.70E-4 |
| 080207 | Swift | 1.58 | 2.09 | 340 | 27.7 | 61 (15-150) | -0.24 | 0.15 | 7.29E-3 | 2.78E-2 |
| 080602 | Swift | 1.43 | 1.82 | 74 | 41.9 | 32 (15-150) | 0.63 | 1.63 | 1.94E-2 | 6.48E-2 |
| 081028A | Swift | 1.25 | 3.04 | 260 | 29.9 | 37 (15-150) | 2.43 | 0.81 | 2.90E-2 | 1.14E-1 |
| 081128 | Swift | 1.98 | <4 | 100 | 31.8 | 23 (15-150) | 0.05 | -0.80 | 1.64E-2 | 1.11E-1 |
| 090407 | Swift | 1.73 | 1.45 | 310 | 45 | 11 (15-150) | -0.78 | 0.16 | 1.58E-2 | 3.86E-2 |
| 090417B | Swift | 1.85 | 0.35 | 260 | 37.2 | 23 (15-150) | -0.52 | 0.12 | 5.91E-3 | 9.93E-4 |
| 090424 | Swift | 1.19 | 0.54 | 48 | 33.1 | 210 (15-150) | -0.15 | -0.06 | 1.56E-3 | 4.47E-4 |
| 090529A | Swift | 2.00 | 2.63 | 100 | 19.9 | 6.8 (15-150) | -0.76 | 0.12 | 2.09E-3 | 1.53E-2 |
| 090902B | Fermi | 1.94 | 1.82 | 19.3 | 23.3 | 4058 (10keV-10GeV) | -0.89 | -1.02 | 5.32E-4 | 1.34E-3 |
| 100206A | Swift | 0.63 | 0.41 | 0.12 | 26.8 | 1.4 (15-150) | -0.71 | 0.85 | 3.80E-5 | 2.49E-5 |
| 100418A | Swift | 2.16 | 0.62 | 7 | 18.7 | 3.4 (15-150) | -0.72 | -0.59 | 1.07E-4 | 2.55E-5 |
| 100424A | Swift | 1.83 | 2.47 | 104 | 33.4 | 15 (15-150) | 1.43 | 1.70 | 1.71E-2 | 8.40E-2 |
| 100728A | Swift | 1.18 | 1.57 | 198.5 | 44.9 | 380 (15-150) | 0.55 | 0.19 | 2.51E-2 | 3.58E-2 |
| 101224A | Swift | 1.05 | 0.72 | 0.2 | 22.6 | 0.58 (15-150) | -0.10 | -0.42 | 4.70E-5 | 2.16E-5 |
| 110106B | Swift | 1.76 | 0.62 | 24.8 | 25.1 | 20 (15-150) | -1.40 | -0.83 | 1.80E-4 | 1.22E-4 |
| 110128A | Swift | 1.31 | 2.34 | 30.7 | 43.2 | 7.2 (15-150) | -0.06 | 0.16 | 1.63E-2 | 5.66E-2 |
| 111211A | AGILE | 2.77 | 0.48 | 15 | 20.3 | 92 (20-1200) | 0.16 | 0.16 | 2.45E-4 | 6.41E-5 |
| 120118B | Swift | 2.08 | 2.94 | 23.2 | 42.7 | 18 (15-150) | -0.71 | -0.45 | 1.53E-2 | 7.77E-2 |
| 120326A | Swift | 2.06 | 1.8 | 69.6 | 41.0 | 26 (15-150) | -0.24 | 0.25 | 1.22E-2 | 3.80E-2 |

| 120716A | IPN | CPL | 2.49 | 230 | 35.7 | 147 (10-1000) | -1.39 | -1.32 | 9.01E-3 | 4.61E-2 |
| 120722A | Swift | 1.90 | 0.96 | 42.4 | 17.7 | 12 (15-150) | -0.18 | -0.80 | 4.20E-4 | 2.36E-4 |
| 120907A | Swift | 1.73 | 0.97 | 16.9 | 40.2 | 6.7 (15-150) | 0.10 | 0.38 | 3.46E-3 | 3.34E-3 |
| 130131B | Swift | 1.15 | 2.54 | 4.3 | 27.3 | 3.4 (15-150) | 0.80 | 0.19 | 1.45E-3 | 5.91E-3 |

**Notes**. Column 1: GRB name (GCN name for Swift GRBs and Trigger ID for Fermi GRBs). Column 2: satellite that detected the burst. Column 3: spectral index reported by the satellite: "CPL" means that the spectrum measured by the satellite is fitted with a cutoff power law. Others are fitted by a simple power law. Column 4: redshift. Column 5: burst duration $T_{90}$ as measured by the respective satellite. Column 6: zenith angle at the ARGO-YBJ location. Column 7: fluence in the keV range measured by the satellite that detected the burst. Column 8 and 9: statistical significance of the cluster detected during the prompt phase, for two $E_{cut}$ values (100 GeV and 1 TeV). Column 10 and 11: 99% C. L. upper limits to the fluence for two $E_{cut}$ values (100 GeV and 1 TeV) with $\alpha = 2.0$ ($\alpha = 1.94$ for GRB 090902B).